\documentclass[11pt, a4paper]{article}
\usepackage[latin9]{inputenc}
\usepackage{amssymb}
\usepackage{eurosym}
\usepackage{amsfonts}
\usepackage{amsmath}
\usepackage{array}
\usepackage{calc}
\usepackage{array}
\usepackage{indentfirst}
\usepackage{graphicx}
\usepackage{hyperref}
\usepackage{authblk}
\usepackage{cite}
\usepackage{color}

\begin{document}

\title{Resonant superalgebras and $\mathcal{N}=1$ supergravity theories in three spacetime dimensions}

\author{Patrick Concha\thanks{%
patrick.concha@ucsc.cl},\thinspace\ Remigiusz Durka\thanks{%
remigiusz.durka@uwr.edu.pl},\thinspace\ Evelyn Rodríguez\thanks{%
evelyn.rodriguez@edu.uai.cl}, \\
\small
$^{\ast}$\textit{Departamento de Matemática y Física Aplicadas,}\\
\textit{Universidad Católica de la Santísima Concepción,}\\
Alonso de Ribera 2850, Concepción, Chile.\\
$^{\dagger}$\textit{Institute for Theoretical Physics, University of Wroc\l{}aw},\\
pl.\ M.\ Borna 9, 50-204 Wroc\l {}aw, Poland\\
$^{\ddag}$\textit{Departamento de Ciencias, Facultad de Artes Liberales,}\\
\textit{Universidad Adolfo Ibáñez}, Viña del Mar, Chile. }
\normalsize
\maketitle

\begin{abstract}
We explore $\mathcal{N}=1$ supersymmetric extensions of algebras going beyond the Poincaré and AdS ones in three spacetime dimensions. Besides reproducing two known examples, we present new superalgebras, which all correspond to supersymmetric extensions with one fermionic charge $Q_{\alpha}$ concerning the so-called resonant algebras being characterized by the presence of an additional bosonic generator $Z_{a}$, besides the Lorentz $J_{a}$ and translation $P_{a}$ generators. Obtained eight supersymmetric $JPZ+Q$ schemes result directly from obeying super-Jacobi identities. We point out particular requirements that superalgebras have to satisfy to be successfully incorporated within valid supergravity actions. The presented algebraic and Lagrangian framework organizes and helps us better understand relations between the various supergravity and supersymmetric Chern-Simons actions invariant under diverse resonant superalgebras.
\end{abstract}

\section{Introduction}

It is well known that three-dimensional supergravity theory \cite{Deser:1982sw} with the presence of cosmological constant can be described by the means of the Chern-Simons (CS) action and AdS superalgebra \cite{Achucarro:1987vz}. At the level of symmetry, the limit of vanishing cosmological constant can be performed through the Inönü-Wigner contraction procedure leading to the Poincaré superalgebra \cite{Achucarro:1989gm}. Subsequently, to transit beyond the three-dimensional $\mathcal{N}=1$ AdS and Poincaré supergravity theories towards more general models and frameworks, various extensions were studied. Among others, this concerns the incorporation of $\mathcal{N}$-extended charges, coupling to matter, higher-spin coupling, non-relativistic and ultra-relativistic limits, and higher-dimensional generalizations (for examples see \cite{Marcus:1983hb, Aragone:1983sz, vanNieuwenhuizen:1985cx, Rocek:1985bk, Howe:1995zm, Banados:1996hi, Giacomini:2006dr, Andringa:2009yc, Andringa:2013mma, Butter:2013rba, Kuzenko:2013uya, Alkac:2014hwa, Fuentealba:2015jma, Fuentealba:2015wza, Bergshoeff:2015ija, Barnich:2015sca, Henneaux:2015tar, Bergshoeff:2016lwr, Basu:2017aqn, Fuentealba:2017fck, Andrianopoli:2019sqe, Ozdemir:2019orp, Ravera:2019ize, Ozdemir:2019tby, Fuentealba:2019bgb, Andrianopoli:2019sip, Ali:2019jjp, Concha:2019mxx, Castellani:2020kmz}). In this article, we will explore another interesting path that leads through the enlarged symmetries. Enlarged algebras equip the set of the generators of Poincaré/AdS algebras by new bosonic generator(s) along with the corresponding gauge fields.

A particular extension and deformation of the Poincaré algebra is given by the so-called Maxwell algebra, which appears to describe a Minkowski spacetime in the presence of a constant electromagnetic background \cite{Schrader:1972zd, Bacry:1970ye, Gomis:2017cmt}. Subsequently, the Maxwell symmetry has been useful to recover standard General Relativity from CS and Born-Infeld (BI) gravity theories \cite{Edelstein:2006se, Izaurieta:2009hz, Concha:2013uhq, Concha:2014vka, Concha:2014zsa}. In three spacetime dimensions, the additional gauge field appearing for the Maxwell algebra modifies not only the vacuum of the theory but also the asymptotic sector \cite{Concha:2018zeb}. More recently, it was shown that the minimal massive three-dimensional gravity \cite{Bergshoeff:2014pca} could be seen as a particular case of a generalized minimal massive gravity theory, which appears as a spontaneous breaking of a local symmetry of the CS gravity invariant under the Maxwell algebra \cite{Chernyavsky:2020fqs} (also known as Hietarinta algebra \cite{Hietarinta:1975fu, Bansal:2018qyz}). Interestingly, the Maxwell (super)algebra can be alternatively obtained as the Inönü-Wigner contraction of the so-called Soroka-Soroka \footnote{Also denoted as AdS-Lorentz algebra or semi-simple enlargement of the Poincaré algebra.} (super)algebra\cite{Soroka:2004fj, Soroka:2006aj}. The latter has been useful to recover diverse Lovelock gravity theories \cite{Concha:2016kdz, Concha:2016tms, Concha:2017nca} from CS and BI theories. Other vast applications of the Maxwell and Soroka-Soroka algebras can be found in \cite{deAzcarraga:2010sw, Durka:2011nf, Durka:2011va, deAzcarraga:2012qj, Diaz:2012zza, Salgado:2014jka, Hoseinzadeh:2014bla, Cebecioglu:2015jta, Palumbo:2016nku, Caroca:2017izc, Aviles:2018jzw, Kibaroglu:2018mcn, Concha:2018jjj, Salgado-Rebolledo:2019kft, Concha:2019lhn, Penafiel:2019czp, Concha:2019eip, Paixao:2019usi, Concha:2020sjt, Adami:2020xkm}.

Both Maxwell and Soroka-Soroka algebras (and other Maxwell-like algebras \cite{Concha:2016hbt}) can be alternatively obtained through the semigroup expansion method \cite{Izaurieta:2006zz, Caroca:2011qs, Andrianopoli:2013ooa, Artebani:2016gwh, Ipinza:2016bfc, Inostroza:2017ezc, Inostroza:2018gzd}. Indeed, they appear as the "resonant" expansions of the AdS algebra \cite{Salgado:2014qqa}. A semigroup expansion is said to be resonant when the decomposition of the semigroup satisfies the same structure as the subspaces of the original (super)algebra. The expansion procedure based on the semigroups has been of particular interest in gravity context since it not only allows us to obtain novel algebras from a known one but also provides us the non-vanishing components of the invariant tensor crucial for the construction of CS and BI actions.

Recently, it was shown in \cite{Durka:2016eun} that the Maxwell and Soroka-Soroka algebras are not the only algebras that can be obtained as the resonant expansion of the AdS algebra. In particular, for the set of generators $J_{a},P_{a},Z_{a}$, the whole family of the resonant algebras consists of 6 cases, denoted as $\mathfrak{B}_{4}$, $\tilde{B}_{4}$, $B_{4}$, $\mathfrak{C}_{4}$, $\tilde{C}_{4}$, $C_{4}$ (with $\mathfrak{C}_4$ and $ \mathfrak{B}_4$ being Soroka-Soroka and Maxwell algebras, respectively). The resonant algebras turned out to be of particular interest for the description of topological insulators \cite{Durka:2019guk}. A detailed study about the number of possible resonant algebras depending on the generator content has then been presented in \cite{Durka:2019vnb}. Now, in this paper, we are going to open similar discussion in the context of the superalgebras.

The supersymmetric extension of the resonant algebras has only been studied for the case of the Maxwell and Soroka-Soroka algebras. This concerns cases with one and two fermionic charges, and with the presence of additional bosonic content \cite{Bonanos:2009wy, Bonanos:2010fw, Lukierski:2010dy, Durka:2011gm, Kamimura:2011mq, deAzcarraga:2012zv, deAzcarraga:2014jpa, Concha:2014xfa, Concha:2014tca, Concha:2015tla, Ipinza:2016con, Penafiel:2017wfr, Ravera:2018vra, Banaudi:2018zmh, Penafiel:2018vpe, Concha:2018ywv, Kibaroglu:2018oue}. Here, we would like to explore the $\mathcal{N}=1$ supersymmetric extension (with one fermionic charge) of the complete set of $JPZ$ resonant algebras. By analyzing the super-Jacobi identities we show that some resonant algebras can present more than one supersymmetric extension scheme. We include the CS actions invariant under all of the respective resonant superalgebras. There are several motivations to do so. On one hand, higher-dimensional supergravity (as its $D=11$ version) requires the presence of an additional bosonic gauge field. Then, study the possible supersymmetric extensions of enlarged symmetry in three spacetime dimensions could lead to interesting toy models sharing many properties with higher-dimensional theories. On the other hand, due to the diverse applications of the resonant algebras and in particular the Maxwell algebra, it seems pertinent to explore their supersymmetric versions. Furthermore, our results could be useful to have a better understanding of the underlying relations between known and new supergravity theories.

The paper is organized as follows: in Section 2, we briefly review the Poincaré and AdS CS supergravity theories in three spacetime dimensions. Section 3 and 4 contain our main results. In Section 3, we present the supersymmetric extensions of the resonant algebras. Beside known Maxwell-like examples, we present the novel superalgebras sharing a Poincaré-like bosonic sub-structure. Additionally, we discuss the requirements for the valid supersymmetric extensions of the resonant algebras. In Section 4, we construct the general super CS actions for all eight resonant superalgebras. We show that among them altogether five supersymmetric extensions are good candidates to construct supergravity actions. Section 5 concludes our work with a discussion about possible future developments.

\section{Poincaré and AdS supergravities}

In this section, we briefly review the AdS superalgebra and its vanishing cosmological constant limit: the Poincaré superalgebra. We also discuss the construction of the three-dimensional CS supergravity actions invariant under these superalgebras. This will help us to highlight crucial ingredients and details needed for the later introduction of the resonant superalgebras.

In three spacetime dimensions, the AdS superalgebra is given by $\mathfrak{osp}\left( 2|1\right) \times \mathfrak{sp}\left( 2\right) $. It is spanned by the set of generators $\left\{ J_{a},P_{a},Q_{\alpha }\right\} $,
which satisfy the following (anti-)commutation relations:
\begin{eqnarray}
\left[ J_{a},J_{b}\right] &=&\epsilon_{abc}\,J^{c}\,,  \notag \\
\left[ J_{a},P_{b}\right] &=&\epsilon _{abc}\,P^{c}\,,  \notag \\
\left[ P_{a},P_{b}\right] &=&\frac{1}{\ell^{2}}\epsilon _{abc}\,J^{c}\,,  \notag \\
\left[ J_{a},Q_{\alpha }\right] &=&\frac{1}{2}\,\left( \gamma _{a}\right)
_{\alpha }^{\text{ }\beta }Q_{\beta }\,, \label{AdS} \\
\left[ P_{a},Q_{\alpha }\right] &=&\frac{1}{2\ell}\,\left( \gamma _{a}\right)
_{\alpha }^{\text{ }\beta }Q_{\beta }\,,\text{ }  \notag \\
\left\{ Q_{\alpha },Q_{\beta }\right\} &=&-\left( \gamma ^{a}C\right)
_{\alpha \beta }P_{a}-\frac{1}{\ell}\left( \gamma ^{a}C\right) _{\alpha \beta }J_{a}\,,
\notag
\end{eqnarray}
where $a,b,c=0,1,2$ are the Lorentz indices raised and lowered with the Minkowski metric $\eta_{ab}$, and $\epsilon_{abc}$ is the Levi-Civita tensor. Here $\alpha =1,2$ are spinorial indices, $C$ is the charge conjugation matrix, and $\gamma^{a}$ are the Dirac matrices in three spacetime dimensions. For later convenience in comparing different algebras, let us present schematically the generator's structure of (\ref{AdS}) as:
	\begin{equation}
	\begin{tabular}[t]{c|cc}
	\lbrack .,.] & J & P \\ \hline
	J & J & P \\
	P & P & J
	\end{tabular}
	\qquad
	\begin{tabular}[t]{c|c}
	\lbrack .,.] & $Q$ \\ \hline
	J & $Q$ \\
	P & $Q $
	\end{tabular}
	\qquad
	\begin{tabular}[t]{c|c}
	\{.,.\} & $Q$ \\ \hline
	$Q$ & $P+J$
	\end{tabular}
	\end{equation}
	
It is interesting to note that there is a subtle redefinition of the generators allowing to rewrite the AdS superalgebra in a different basis. Indeed, such superalgebra after following the redefinition of the generators
\begin{equation}
J_{a} = \tilde{J}_{a}+\tilde{M}_{a}\,, \qquad P_{a} = \frac{1}{\ell}\left( \tilde{J}_{a}-\tilde{M}
_{a}\right)\,, \qquad Q_{\alpha } = \sqrt{\frac{2}{\ell}}\tilde{Q}_{\alpha }\,.
\end{equation}
can be written as a direct sum of the super Lorentz spanned by $\left\{ \tilde{J}_{a},\tilde{Q}_{\alpha }\right\} $ and the Lorentz algebra spanned by $\left\{ \tilde{M}_{a}\right\} $
\begin{eqnarray}
\left[ \tilde{J}_{a},\tilde{J}_{b}\right] &=&\epsilon _{abc}\tilde{J}^{c}\,,
\notag \\
\left[ \tilde{J}_{a},\tilde{Q}_{\alpha }\right] &=&\frac{1}{2}\,\left(
\gamma _{a}\right) _{\alpha }^{\text{ }\beta }\tilde{Q}_{\beta }\,,  \notag
\\
\text{ \ }\left[ \tilde{M}_{a},\tilde{M}_{b}\right] &=&\epsilon _{abc}\tilde{M}^{c}\,,  \label{SLL} \\
\left\{ \tilde{Q}_{\alpha },\tilde{Q}_{\beta }\right\} &=&-\left( \gamma
^{a}C\right) _{\alpha \beta }\tilde{J}_{a}\,. \notag
\end{eqnarray}
The CS action based on the AdS superalgebra written as (\ref{SLL}) is given by two exotic CS terms \cite{Witten:1988hc}, where only one is supersymmetric. Additionally, although the basis $\left\{\tilde{J}_{a},\tilde{M}_{a},\tilde{Q}_{\alpha}\right\}$ seems simpler, the vanishing cosmological constant limit leading to the Poincaré superalgebra can be done in the basis $\left\{ J_{a},P_{a},Q_{\alpha }\right\}$. Indeed, considering  in (\ref{AdS}) the limit $\ell \rightarrow \infty $ effectively gives the expected superalgebra:
\begin{equation}
\begin{tabular}[t]{c|cc}
\lbrack .,.] & J & P \\ \hline
J & J & P \\
P & P & 0%
\end{tabular}
\qquad
\begin{tabular}[t]{c|c}
\lbrack .,.] & $Q$ \\ \hline
J & $Q$ \\
P & $0 $
\end{tabular}
\qquad
\begin{tabular}[t]{c|c}
\{.,.\} & $Q$ \\ \hline
$Q$ & $P$
\end{tabular}
\end{equation}
The AdS supergravity action is straightforwardly obtained from the CS model defined in three spacetime dimensions by
\begin{equation}
I_{CS}=\frac{k}{4\pi }\int_{\mathcal{M}}\left\langle A \wedge dA+\frac{1}{3} A\wedge [A,A]\right\rangle \,,  \label{CS}
\end{equation}
where $k=1/4G$ is the CS\ level of the theory related to the
gravitational constant $G$. The gauge field $A$ is the gauge connection one-form and $\left\langle \dots \right\rangle $ denotes the invariant trace.

The gauge-connection for the AdS superalgebra reads
\begin{equation}
A=\omega ^{a}J_{a}+ e^{a}P_{a}+ \psi ^{\alpha }Q_{\alpha }\,,  \label{gc}
\end{equation}
where $\omega ^{a}$ is the spin-connection, $e^{a}$ denotes the vielbein and $\psi ^{\alpha }$ corresponds to the gravitino one-form. The corresponding full curvature two-form $F=dA+\frac{1}{2}\left[ A,A\right] $ is given by
\begin{equation}
F=R^{a}J_{a}+ T^{a}P_{a}+\mathcal{F}^{\alpha}Q_{\alpha }\,,
\end{equation}%
with the super-Lorentz, super-torsion, and fermionic curvatures defined as:
\begin{eqnarray}
R^{a} &=&\mathcal{R}^{a}+\frac{1}{\ell^2} e^{a}e^{b}+\frac{1}{2\ell}\bar{\psi}\gamma ^{a}\psi \,,
\notag \\
T^{a} &=&D_{\omega }e^{a}+\frac{1}{2}\bar{\psi}\gamma ^{a}\psi \,, \\
\mathcal{F} &=&\mathcal{D}_{\omega }\psi +\frac{1}{2\ell}e^{a}\gamma _{a}\psi \,.  \notag
\end{eqnarray}
In above, we have omitted the wedge product $\wedge $ and defined the Lorentz covariant derivative as $D_{\omega }=d+\omega $. Note, that $\mathcal{R}^{a}=d\omega ^{a}+\frac{1}{2}\epsilon ^{abc}\omega _{b}\omega _{c}$ denotes the Lorentz curvature two-form, $D_{\omega }e^{a}=de^{a}+\epsilon^{abc}\omega_{b}e_{c}$ defines the torsion two-form, whereas in the fermionic curvature we have $\mathcal{D}_{\omega }\psi=d\psi+\frac{1}{2}\omega^{a} \gamma_{a} \psi$.

The AdS superalgebra (\ref{AdS}) admits the following non-vanishing components of the invariant tensor,
\begin{eqnarray}
\left\langle J_{a}J_{b}\right\rangle &=&\alpha _{0}\,\eta _{ab}\,,  \notag \\
\left\langle J_{a}P_{b}\right\rangle &=&\frac{\alpha _{1}}{\ell}\,\eta _{ab}\,,  \notag \\
\left\langle P_{a}P_{b}\right\rangle &=&\frac{\alpha _{0}}{\ell^{2}}\,\eta _{ab}\,,
\label{adsinvt} \\
\left\langle Q_{\alpha }Q_{\beta }\right\rangle &=&\frac{2}{\ell}\left( \alpha
_{1}-\alpha _{0}\right) C_{\alpha \beta }\,,  \notag
\end{eqnarray}%
where $\alpha _{0}$ and $\alpha _{1}$ are arbitrary constants. In order to define a proper flat limit at the level of the invariant tensor and the CS action, let us consider the following redefinition of the coupling
constants,
\begin{equation}
\alpha_{0}\rightarrow\alpha_{0}\,,\qquad  \alpha_{1}\rightarrow\ell\alpha_{1}\,.
\end{equation}

With this redefinition, the components of the invariant tensor (\ref{adsinvt}) lead in the vanishing cosmological constant limit $\ell \rightarrow \infty $, to those of the Poincaré. Through considering superalgebra (\ref{AdS}) with its invariant tensors (\ref{adsinvt}) (along with the above redefinition) and the respective gauge connection one-form (\ref{gc}) in the definition of the CS action (\ref{CS}) with $k=\frac{1}{4G}$, we find the most general three-dimensional $\mathcal{N}=1$ AdS CS supergravity action \cite{Giacomini:2006dr} to be of the form:
\begin{eqnarray}
I_{CS}^{AdS} &=&\frac{k}{4\pi }\int \left[ \alpha _{0}\left( \omega
^{a}d\omega _{a}+\frac{1}{3}\epsilon ^{abc}\omega _{a}\omega _{b}\omega
_{c}+\frac{1}{\ell^2}e_{a}D_{\omega }e^{a}-\frac{2}{\ell}\bar{\psi}\mathcal{F} \right) \right.  \notag \\
&&\left. +\alpha _{1}\left( 2\mathcal{R}^{a}e_{a}+\frac{1}{3\ell^2}\epsilon^{abc}e_{a}e_{b}e_{c}+2\bar{\psi}\mathcal{F} \right) \right] \,.
\end{eqnarray}
One can see that the AdS CS supergravity action contains two independent terms proportional to $\alpha _{0}$ and $\alpha_{1}$. In particular, the $\alpha_{0}$ term contains the so-called exotic Lagrangian \cite{Witten:1988hc}, along with the torsional term plus a contribution from the gravitino. In particular, the bosonic terms in the first bracket are related to the Pontryagin and  Nieh-Yan forms \cite{Troncoso:1999pk}. The usual three-dimensional supergravity Lagrangian with a cosmological constant is proportional to the $\alpha _{1}$ term whose bosonic part, given by the Einstein-Hilbert term, is related to the Euler density. After considering the limit $\ell \rightarrow \infty $, the resulting CS action
\begin{equation}
I_{CS}^{Poincare}=\frac{1}{16\pi G }\int \left[ \alpha_{0}\left( \omega
^{a}d\omega _{a}+\frac{1}{3}\epsilon ^{abc}\omega _{a}\omega _{b}\omega
_{c} \right) +\alpha _{1}\left( 2\mathcal{R}^{a}e_{a}+2\bar{\psi} \mathcal{D}_{\omega}\psi \right) \right] \,.
\end{equation}
describes the three-dimensional $\mathcal{N}=1$ Poincaré CS supergravity.

\section{Supersymmetric extensions of the resonant algebras}

In this section, we shall discuss the supersymmetric extensions of the resonant bosonic algebras. Such algebras were obtained through the semigroup expansion method \cite{Izaurieta:2006zz} based on the underlying structure of the AdS algebra. Leaving all the details aside, effectively they all can be seen as the enlargement of the set of generators, Lorentz and translations, including additional bosonic generators.

For just two generators, $J_{a}$ and $P_{a}$, we can properly close an algebra by the means of two scenarios given by the Poincaré and AdS algebras. Adding the fermionic generator $Q_{\alpha}$ leads to their two $JP+Q$ supersymmetric versions. As it was shown in \cite{Durka:2016eun, Durka:2019guk, Durka:2019vnb}, the set of $\{J_{a},P_{a},Z_{a}\}$ leads to six different ways we can close an algebra, which we denote as $\mathfrak{B}_{4}$, $\mathfrak{C}_{4}$, $B_{4}$, $C_{4}$, $\tilde{B}_{4}$, $\tilde{C}_{4}$. Supersymmetric extensions of the $\mathfrak{B}_{4}$ and $\mathfrak{C}_{4}$, known as Maxwell and Soroka-Soroka algebras respectively, have already been explored in wide range of spacetime dimensions for various purposes \cite{Soroka:2006aj, Bonanos:2009wy, Bonanos:2010fw, Lukierski:2010dy, Durka:2011gm, Kamimura:2011mq, deAzcarraga:2012zv, deAzcarraga:2014jpa, Concha:2014xfa, Concha:2014tca, Concha:2015tla, Ipinza:2016con, Penafiel:2017wfr,
Ravera:2018vra, Banaudi:2018zmh, Penafiel:2018vpe, Concha:2018ywv, Kibaroglu:2018oue}. Generalizations to infinite-dimensional enhancements of the Maxwell and Soroka-Soroka superalgebras, as well as the incorporation
of central charges, have also been studied in \cite{Caroca:2019dds}. In three spacetime dimensions, supersymmetric theories based on the Maxwell and Soroka-Soroka superalgebras have been constructed in \cite{Fierro:2014lka, Concha:2015woa, Concha:2016zdb, Concha:2018jxx, Concha:2019icz}. Although the fermionic generator $Q_{\alpha}$ being present in the superalgebra leads to the supersymmetric CS model, this not necessarily assures the supergravity action. For instance, the Maxwell case provides such a non-standard supersymmetric extension \cite{Lukierski:2010dy} as we lack the translation generator $P_{a}$ in the outcome of the anti-commutator of $\left\{Q,Q\right\}\neq 0$. To achieve $\left\{Q,Q\right\} \sim P$, being obligatory for supergravity as this provides coupling of the gravitino term with the Einstein-Hilbert term describing graviton, it was necessary to include yet additional fermionic charge $\Sigma_{\alpha}$. Such a scenario was called minimal supersymmetric extension. Both types of the $\mathfrak{C}_{4}$ superalgebras, with single and two fermionic charges, allow us to define supergravity CS actions but only the minimal one for the vanishing cosmological constant leads to an invariant CS supergravity action under the minimal Maxwell superalgebra \cite{Concha:2018jxx, Concha:2019icz}.

Now with more algebras coming from $JPZ$ configurations, this paper attempts to answer the question of how many valid supersymmetric versions of resonant algebras we can find for the generator content $JPZ+Q$. To find all the possible realizations, we must first ask which features are necessary in order to get a valid superalgebra, and what is required to successfully build a gauge theory of supergravity.

The minimal requirement for the supersymmetric extension of the resonant algebra should be possessing a Lorentz subalgebra spanned by $J$. Then, every bosonic and fermionic generator given by $B$ and $F$, respectively, has to be invariant under Lorentz rotations $J$,
\begin{equation}
\left[ J,B\right] \sim B\,,\qquad \left[ J,F\right] \sim F\,.
\end{equation}

Additionally, we must include the so-called resonant condition, which organizes the outcomes of the commutation relations for the \textit{L=Lorentz-like} and \textit{T=translational-like} generators to assure a structure incorporated within the original AdS:
\begin{equation*}
\left[ L,L \right] \sim L\,,\qquad \left[ L,T\right] \sim T\,,\qquad \left[T,T\right] \sim L\,.
\end{equation*}
This extends also for the fermions: $\left[ L,Q\right] \sim Q$ and $\left[T,Q\right]\sim Q$. Although it is not visible on the first sight, a new generator $Z$ presented in this paper is of the Lorentzian-like type. This is due to the fact that the particular $Z$ generator appearing in the resonant algebras has been obtained as an expansion of the Lorentz generator $J$ \cite{Durka:2016eun, Durka:2019guk, Durka:2019vnb}.

On the other hand, in order to have true superalgebras we need that the bosonic and fermionic generators satisfy
\begin{equation}
\left[ B,B\right] \sim B\,,\qquad \left[ B,F\right] \sim F\,,\qquad \left\{
F,F\right\} \sim B\,.
\end{equation}
Ultimately, we must require that all the generators satisfy the super-Jacobi identities. Naturally, we shall avoid candidates possessing relation $\left\{ Q,Q\right\}=0$ passing Jacobi identities as well as other requirements above, since they would imply the absence of fermionic contributions to the bosonic curvatures, and in turn, to the action. Note that the AdS and Poincaré supergravities are characterized by the anticommutator $\{Q,Q\}\sim P$. The generator $P_{a}$ is there expressed as bilinear expressions of fermionic generators $Q$ to assure the supersymmetric action describing a supergravity theory coupling the graviton (Einstein-Hilbert) with the gravitino (Rarita-Schwinger). We are going to allow a much wider scope of superalgebras allowing the fermionic contributions in other sectors of the CS action different to the Einstein-Hilbert one. In that non-standard case, we would describe such action not as supergravity but as the supersymmetric Chern-Simons theory.

Extensive searches of all possibilities, aided with the \textit{Cadabra} symbolic computer algebra system \cite{PeetersK1,PeetersK2} have resulted ultimately in eight superalgebras, among them five leads to the supergravity formulation and three to the non-standard cases. Two represent already known non-standard supersymmetric extension of $\mathfrak{C}_{4}$ and $\mathfrak{B}_{4}$ algebras, Soroka-Soroka \cite{Soroka:2006aj} and Maxwell \cite{Lukierski:2010dy}, respectively:
\begin{equation*}
\begin{array}{llllllll}
\mathfrak{C}_{4}: &  &
\begin{tabular}[t]{c|ccc}
\lbrack .,.] & J & P & Z \\ \hline
J & J & P & Z \\
P & P & Z & P \\
Z & Z & P & Z
\end{tabular}
&  &
\begin{tabular}[t]{c|c}
\lbrack .,.] & Q \\ \hline
J & Q \\
P & Q \\
Z & Q
\end{tabular}
&  &
\begin{tabular}[t]{c|c}
\{.,.\} & Q \\ \hline
Q & P+Z
\end{tabular}
& ~~~\\[55pt]
\mathfrak{B}_{4}: &  &
\begin{tabular}[t]{c|ccc}
\lbrack .,.] & J & P & Z \\ \hline
J & J & P & Z \\
P & P & Z & 0 \\
Z & Z & 0 & 0
\end{tabular}
&  &
\begin{tabular}[t]{c|c}
\lbrack .,.] & Q \\ \hline
J & Q \\
P & 0 \\
Z & 0
\end{tabular}
&  &
\begin{tabular}[t]{c|c}
\{.,.\} & Q \\ \hline
Q & Z
\end{tabular}
& ~~~
\end{array}
\end{equation*}
On the other hand, all novel supersymmetric extensions of the resonant algebras correspond to $C_{4},\tilde{C}_{4},\tilde{B}_{4}$ cases, which share the Poincaré-like sub-structure \cite{Durka:2016eun, Durka:2019guk, Durka:2019vnb}:
\begin{equation*}
\begin{array}{llllllll}
C_{4}: &  &
\begin{tabular}[t]{c|ccc}
\lbrack .,.] & J & P & Z \\ \hline
J & J & P & Z \\
P & P & 0 & P \\
Z & Z & P & Z
\end{tabular}
&  &
\begin{tabular}[t]{c|c}
\lbrack .,.] & Q \\ \hline
J & Q \\
P & 0 \\
Z & Q
\end{tabular}
&  &
\begin{tabular}[t]{c|c}
\{.,.\} & Q \\ \hline
Q & P
\end{tabular}
&  \\[55pt]
\tilde{C}_{4}: &  &
\begin{tabular}[t]{c|ccc}
\lbrack .,.] & J & P & Z \\ \hline
J & J & P & Z \\
P & P & 0 & 0 \\
Z & Z & 0 & Z%
\end{tabular}
&  &
\begin{tabular}[t]{c|c}
\lbrack .,.] & Q \\ \hline
J & Q \\
P & 0 \\
Z & 0
\end{tabular}
&  &
\begin{tabular}[t]{c|c}
\{.,.\} & Q \\ \hline
Q & P
\end{tabular}
& $\textcircled{\footnotesize 1}$ \\[55pt]
&  &  &  &
\begin{tabular}[t]{c|c}
\lbrack .,.] & Q \\ \hline
J & Q \\
P & 0 \\
Z & Q
\end{tabular}
&  &
\begin{tabular}[t]{c|c}
\{.,.\} & Q \\ \hline
Q & Z
\end{tabular}
& $\textcircled{\footnotesize 2}$ \\[55pt]
\tilde{B}_{4}: &  &
\begin{tabular}[t]{c|ccc}
\lbrack .,.] & J & P & Z \\ \hline
J & J & P & Z \\
P & P & 0 & 0 \\
Z & Z & 0 & 0
\end{tabular}
&  &
\begin{tabular}[t]{c|c}
\lbrack .,.] & Q \\ \hline
J & Q \\
P & 0 \\
Z & 0
\end{tabular}
&  &
\begin{tabular}[t]{c|c}
\{.,.\} & Q \\ \hline
Q & P
\end{tabular}
& $\textcircled{\footnotesize 1}$ \\[25pt]
&  &  &  &  &  &
\begin{tabular}[t]{c|c}
\{.,.\} & Q \\ \hline
Q & Z
\end{tabular}
& $\textcircled{\footnotesize 2}$ \\[25pt]
&  &  &  &  &  &  &  \\
&  &  &  &  &  &
\begin{tabular}[t]{c|c}
\{.,.\} & Q \\ \hline
Q & P+Z%
\end{tabular}
& $\textcircled{\footnotesize 3}$ \\
&  &  &  &  &  &
\end{array}
\end{equation*}
It should be pointed out that one can not construct a consistent supersymmetric extension with one spinor charge for the $B_{4}$ algebra:
\begin{equation*}
\begin{array}{llllllll}
B_{4}: &  &
\begin{tabular}[t]{c|ccc}
\lbrack .,.] & J & P & Z \\ \hline
J & J & P & Z \\
P & P & 0 & P \\
Z & Z & P & 0%
\end{tabular}
&  &
\begin{tabular}[t]{c|c}
\lbrack .,.] & Q \\ \hline
J & - \\
P & - \\
Z & -
\end{tabular}
&  &
\begin{tabular}[t]{c|c}
\{.,.\} & Q \\ \hline
Q & -%
\end{tabular}
& ~~~~~~
\end{array}
\end{equation*}

As the Soroka-Soroka $\mathfrak{C}_{4}$ superalgebra represents the most general form without any zero entries, it is good to complete this section with its explicit form in three spacetime dimensions:
\begin{eqnarray}
\left[ J_{a},J_{b}\right] &=&\epsilon _{abc}J^{c}\,,\qquad \ \ \ \left[
J_{a},P_{b}\right] =\epsilon _{abc}P^{c}\,,  \notag \\
\left[ P_{a},P_{b}\right] &=&\epsilon _{abc}Z^{c}\,,\qquad \ \ \ \left[
J_{a},Z_{b}\right] =\epsilon _{abc}Z^{c}\,, \notag \\
\left[ Z_{a},Z_{b}\right] &=&\frac{1}{\ell^2}\epsilon _{abc}Z^{c}\,,\qquad \left[
P_{a},Z_{b}\right] =\frac{1}{\ell^2}\epsilon _{abc}P^{c}\,,  \notag
\end{eqnarray}
\begin{eqnarray}
\left[ J_{a},Q_{\alpha }\right] &=&\frac{1}{2}\,\left( \gamma _{a}\right)
_{\alpha }^{\text{ }\beta }Q_{\beta }\,,  \label{SorokaExplicit}\\
\left[ P_{a},Q_{\alpha }\right] &=&\frac{1}{2 \ell}\,\left( \gamma _{a}\right)
_{\alpha }^{\text{ }\beta }Q_{\beta }\,, \notag \\
\left[ Z_{a},Q_{\alpha }\right] &=&\frac{1}{2 \ell^2}\,\left( \gamma _{a}\right)
_{\alpha }^{\text{ }\beta }Q_{\beta }\,,  \notag
\end{eqnarray}
\begin{eqnarray}
\left\{ Q_{\alpha },Q_{\beta }\right\} &=&-\left( \gamma ^{a}C\right)
_{\alpha \beta }(\frac{1}{\ell}P_{a}+Z_{a})\,. \notag
\end{eqnarray}
This allows us to easily reproduce the explicit form for any other scheme presented in the $JPZ+Q$ tables given earlier. All other superalgebras will represent merely alterations in existence of the particular outcomes.

It is important to clarify that the supersymmetric extensions of the resonant algebras presented here have not been obtained through the semigroup expansion method. Therefore, they are not necessarily resonant expansions of the AdS superalgebra. Obtained results were the effect of the brute-force analysis of the super-Jacobi identities, with several well-motivated restrictions on behalf of the (anti-)commutators. The resonant property formally appears on their bosonic subalgebras, which correspond to the respective resonant algebras. The exploration of superalgebras being resonant expansions of a known superalgebra remains an interesting open issue. In that respect, it would be more appropriate to approach it in the "minimal" scheme, in which minimal supersymmetric extensions of $\mathfrak{C}_{4}$ and $\mathfrak{B}_{4}$ can be obtained as a resonant expansion of the Lorentz superalgebra \cite{Concha:2018jxx, Concha:2019icz}.

\section{Supergravity and supersymmetric actions based on resonant
superalgebras}

In this section, with the given supersymmetric extensions of the resonant algebras, we will construct the corresponding supersymmetric CS actions in three spacetime dimensions. Analogously to \cite{Durka:2019guk}, we will start from the explicit $\mathfrak{C}_{4}$ superalgebra and the corresponding CS action construction, as it provides the richest Lagrangian content. For other superalgebras, we luckily avoid reshuffling terms into various sub-invariant sectors, so it is all reducing down to either existence of the particular term in the action or its absence.

Let us consider first the gauge connection one-form
\begin{equation}
A=\omega ^{a}J_{a}+e^{a}P_{a}+h^{a}Z_{a}+\psi ^{\alpha }Q_{\alpha }\,,
\label{C41f}
\end{equation}
where $\omega ^{a}$ is the spin-connection one-form, $e^{a}$ is the vielbein, $h^{a}$ is the additional gauge field related to $Z_{a}$ and $\psi $ is the gravitino.

The corresponding super-curvature two-form $F=dA+\frac{1}{2}\left[ A,A\right] $ reads
\begin{equation}
F=R^{a}J_{a}+T^{a}P_{a}+H^{a}Z_{a}+\mathcal{F}^{\alpha }Q_{\alpha }\,, \label{curvs}
\end{equation}
where
\begin{align}
R^{a}& =\mathcal{R}^{a}\,,  \notag \\
T^{a}& =D_{\omega }e^{a}+\frac{1}{\ell^2}\epsilon^{abc}h_{b}e_{c}+\frac{1}{2 \ell}\bar{\psi}\gamma ^{a}\psi \,,  \notag \\
H^{a}& =D_{\omega }h^{a}+\frac{1}{2\ell^2}\epsilon ^{abc}h_{b}h_{c}+\frac{1}{2}\epsilon ^{abc}e_{b}e_{c}+\frac{1}{2\ell}\bar{\psi}\gamma ^{a}\psi \,,  \notag \\
\mathcal{F} & =\mathcal{D}_{\omega }\psi +\frac{1}{2\ell}e^{a}\gamma _{a}\psi +\frac{1}{2\ell^2}h^{a}\gamma _{a}\psi \,,  \notag
\end{align}

The Soroka-Soroka $\mathfrak{C}_{4}$ superalgebra admits the following
non-vanishing components of the non-degenerate invariant tensor \cite{Fierro:2014lka},
\begin{eqnarray}
\left\langle J_{a}J_{b}\right\rangle &=&\alpha _{0}\eta _{ab}\,,\qquad
\left\langle P_{a}P_{b}\right\rangle =\frac{\alpha _{2}}{\ell^2}\eta _{ab}\,,  \notag \\
\left\langle J_{a}P_{b}\right\rangle &=&\frac{\alpha _{1}}{\ell}\eta _{ab}\,,\qquad
\left\langle J_{a}Z_{b}\right\rangle =\frac{\alpha _{2}}{\ell^2}\eta _{ab}\,,  \notag \\
\left\langle Z_{a}P_{b}\right\rangle &=&\frac{\alpha _{1}}{\ell^3}\eta _{ab}\,,\qquad
\left\langle Z_{a}Z_{b}\right\rangle =\frac{\alpha _{2}}{\ell^4}\eta _{ab}\,,  \label{C4inv}
\\
\left\langle Q_{\alpha }Q_{\beta }\right\rangle &=&\frac{2}{\ell^{2}}\left( \alpha
_{1}-\alpha _{2}\right) C_{\alpha \beta }\,,  \notag
\end{eqnarray}%
where $\alpha _{0}$, $\alpha _{1}$ and $\alpha _{2}$ are arbitrary constants. Note that their appearance for different superalgebras is also directly associated with the form of particular $JPZ+Q$ tables \cite{Durka:2016eun,Durka:2019guk, Durka:2019vnb}. Particular set of the $\alpha's$ constants can be seen as the result of the semigroup expansion procedure applied to the original (super)algebra. In fact, one of the advantage of working with the semigroup expansion method is that it provides automatically the invariant tensors of the expanded (super)algebra. For instance, in the case of the Soroka-Soroka superalgebra, the arbitrary constants appearing in (\ref{C4inv}) are coming from the original AdS supertrace \cite{Diaz:2012zza}. Obviously, is not always possible to relate two (super)algebras through an expansion. One could find the non-vanishing components of the invariant tensor for a given (super)algebra, by just considering the realization of the identity $\langle[X_A , X_B ]\, X_C \rangle = \langle X_A \, [X_B , X_C ]  \rangle$, where $X_A$ are the generators of the (super)algebra.

Analogously to the AdS case, we consider the following redefinition of the arbitrary constants,
\begin{equation}
\alpha_{0}\rightarrow\alpha_{0}\,,\qquad \alpha_{1}\rightarrow\ell\alpha_{1}\,,\qquad
\alpha_{2}\rightarrow\ell^{2}\alpha_{2}\,,\label{redefal}
\end{equation}
to have a right flat limit at the level of invariant tensors and CS action. Indeed, the non-vanishing components of the invariant tensor for the Maxwell algebra are directly obtained from (\ref{C4inv}) after the limit $\ell\rightarrow\infty$.
Then, considering the non-vanishing components of the invariant tensor for the $\mathfrak{C}_{4}$ superalgebra (\ref{C4inv}) (along with the redefinition (\ref{redefal})) and the respective gauge connection one-form (\ref{C41f}) in the general expression of the CS action (\ref{CS}), we find the most general three-dimensional $\mathcal{N}=1$ CS supergravity action (up to the boundary terms) being invariant under the Soroka-Soroka $\mathfrak{C}_{4}$ superalgebra:
\begin{eqnarray}
I_{CS}^{\mathfrak{C}_{4}} &=&\frac{k}{4\pi }\int \left[ \alpha _{0}\left(
\omega ^{a}d\omega _{a}+\frac{1}{3}\epsilon ^{abc}\omega _{a}\omega
_{b}\omega _{c}\right) \right.  \notag \\
&&\left. +\alpha _{1}\left( 2\mathcal{R}^{a}e_{a}+\frac{1}{3\ell^2}\epsilon
^{abc}e_{a}e_{b}e_{c}+\frac{2}{\ell^2}e_{a}D_{\omega }h^{a}+\frac{1}{\ell^4}\epsilon ^{abc}e_{a}h_{b}h_{c}+\frac{2}{\ell}
\bar{\psi}\mathcal{F} \right) \right.  \label{SSCS} \\
&&\left. +\alpha _{2}\left(e_{a}D_{\omega }e^{a}+2h_{a}\mathcal{R}
^{a}+\frac{1}{\ell^2}\epsilon^{abc}e_{a}e_{b}h_{c}+\frac{1}{\ell^2}h_{a}D_{\omega }h^{a}+\frac{1}{3 \ell^4}\epsilon ^{abc}h_{a}h_{b}h_{c}-2\bar{\psi}\mathcal{F} \right) \right] \,.  \notag
\end{eqnarray}%
One can see that the CS supergravity action contains three independent sectors proportional to $\alpha _{0}$, $\alpha _{1}$ and $\alpha _{2}$. In particular, unlike the AdS case, the $\alpha _{0}$ term does not contain a fermionic term and reduces to the exotic Lorentz Lagrangian $\mathcal{L}^{exotic}=CS(\omega)$ \cite{Witten:1988hc}. The term proportional to $\alpha _{1}$ contains the Einstein-Hilbert term, the cosmological constant term, a generalized Rarita-Schwinger term with the presence of the additional gauge field $h^{a}$ in curvature $\mathcal{F}$ and the coupling of the extra gauge field with the vielbein and spin-connection. A new contribution is proportional to $\alpha _{2}$ and contains a torsional term, a generalized cosmological constant term, and a generalized Rarita-Schwinger term.

It is interesting to note that a vanishing cosmological constant limit leads to the following non-standard supersymmetric CS action,
\begin{eqnarray}
I_{CS}^{\mathfrak{B}_{4}} &=&\frac{k}{4\pi }\int \left[ \alpha _{0}\left(
\omega ^{a}d\omega _{a}+\frac{1}{3}\epsilon ^{abc}\omega _{a}\omega
_{b}\omega _{c}\right) \right.  \notag \\
&&\left. +\alpha _{1}\left( 2\mathcal{R}^{a}e_{a}\right) +\alpha _{2}\left(
e_{a}D_{\omega }e^{a}+2h_{a}\mathcal{R}^{a}-2\bar{\psi}\mathcal{D}_{\omega }\psi\right) \right]
\,.
\end{eqnarray}
Although such CS action contains the fermionic gauge field, one can see that the term proportional to $\alpha _{1}$ reduces only to the Einstein-Hilbert term and does not describe a proper supergravity Lagrangian. This supersymmetric CS action is invariant under the non-standard supersymmetric extension of the Maxwell algebra $\mathfrak{B}_{4}$.

Since the CS supergravity action for the $\mathfrak{C}_{4}$ superalgebra (\ref{SSCS}) contains all possible terms, the other supersymmetric CS actions will be uniquely characterized by the absence of particular terms. The following table summarizes all the diverse supersymmetric CS actions constructed from the $JPZ+Q$ resonant superalgebras:
\begin{equation*}
		\scalebox{1}{
	\begingroup
	\setlength{\tabcolsep}{8pt} 
	\renewcommand{\arraystretch}{1.1} 
\begin{tabular}{l||c|c|c|c|c|c|c|c|c}
& $\tilde{B}_{4}^{\textcircled{\footnotesize 1}}$ & $\tilde{B}_{4}^{%
	\textcircled{\footnotesize 2}}$ & $\tilde{B}_{4}^{\textcircled{\footnotesize
		3}}$ & $\tilde{C}_{4}^{\textcircled{\footnotesize 1}}$ & $\tilde{C}_{4}^{%
	\textcircled{\footnotesize 2}}$ & $B_{4}$ & $C_{4}$ & $\mathfrak{B}_{4}$ & $%
\mathfrak{C}_{4}$ \\ \hline\hline
$2 e_{a}\mathcal{R}^{a}$ & $\alpha _{1}$ & $\alpha _{1}$ & $\alpha _{1}$ & $%
\alpha _{1}$ & $\alpha _{1}$ & $\alpha _{1}$ & $\alpha _{1}$ & $\alpha _{1}$
& $\alpha _{1}$ \\
$\frac{2}{\ell^2}e_{a}D_{\omega }h^{a}$ & $\alpha _{1}$ & $\alpha _{1}$ & $\alpha _{1}$ & $ $ & $ $ & $ $ & $\alpha _{1}$ & $ $ & $\alpha _{1}$ \\
$\frac{1}{\ell^4}\epsilon ^{abc}e_{a}h_{b}h_{c}$ & $ $ & $ $ & $ $ & $ $ & $ $ & $ $ & $%
\alpha _{1}$ & $ $ & $\alpha _{1}$ \\
$\frac{1}{3\ell^2}\epsilon ^{abc}e_{a}e_{b}e_{c}$ & $ $ & $ $ & $ $ & $ $ & $ $ & $ $ & $ $ & $ $ & $\alpha _{1}$ \\ \hline
$CS(\omega)$ & $\alpha _{0}$ & $\alpha _{0}$ & $\alpha _{0}$ & $%
\alpha _{0}$ & $\alpha _{0}$ & $\alpha _{0}$ & $\alpha _{0}$ & $\alpha _{0}$
& $\alpha _{0}$ \\
$2 h_{a}\mathcal{R}^{a}$ & $\alpha _{2}$ & $\alpha _{2}$ & $\alpha _{2}$ & $%
\alpha _{2}$ & $\alpha _{2}$ & $\alpha _{2}$ & $\alpha _{2}$ & $\alpha _{2}$
& $\alpha _{2}$ \\
$\frac{1}{\ell^2}h_{a}D_{\omega }h^{a}$ & $ $ & $ $ & $ $ & $\alpha _{2}$ & $\alpha _{2}$ & $ $ & $\alpha _{2}$ & $ $ & $\alpha _{2}$ \\
$\frac{1}{3\ell^4}\epsilon ^{abc}h_{a}h_{b}h_{c}$ & $ $ & $ $ & $ $ & $\alpha _{2}$
& $\alpha _{2}$ & $ $ & $\alpha _{2}$ & $ $ & $\alpha _{2}$ \\
$\frac{1}{\ell^2}\epsilon ^{abc}e_{a}e_{b}h_{c}$ & $ $ & $ $ & $ $ & $ $ & $ $ & $ $ & $ $ &
$ $ & $\alpha _{2}$ \\
$e_{a}D_{\omega }e^{a}$ & $ $ & $ $ & $ $ & $ $ & $ $ & $ $ & $ $ & $\alpha
_{2}$ & $\alpha _{2}$ \\ \hline
$2\bar{\psi}\mathcal{D}_{\omega }\psi $ & $\frac{\alpha _{1}}{\ell}$ & $-\alpha _{2}$ & $\frac{\alpha _{1}}{\ell}-\alpha _{2}$ & $\frac{\alpha _{1}}{\ell}$ & $-\alpha _{2}$ & $ $ & $\frac{\alpha _{1}}{\ell}$ & $%
-\alpha _{2}$ & $\frac{\alpha _{1}}{\ell}-\alpha _{2}$ \\
$\frac{1}{\ell}\bar{\psi}e^{a}\gamma _{a}\psi $ & $ $ & $ $ & $ $ & $ $ & $ $ & $ $ & $ $ & $ $ & $%
\frac{\alpha _{1}}{\ell}-\alpha _{2}$ \\
$\frac{1}{\ell^2}\bar{\psi}h^{a}\gamma _{a}\psi $ & $ $ & $ $ & $ $ & $ $ & $-\alpha _{2}$ & $ $ & $%
\frac{\alpha _{1}}{\ell}$ & $ $ & $\frac{\alpha _{1}}{\ell}-\alpha _{2}$%
\end{tabular}
\endgroup
}
\end{equation*}

It is interesting to note that five superalgebras produce the Rarita-Schwinger term in the same sector as the Einstein-Hilbert term, which describes consistent supergravity. Two superalgebras include the additional contribution from the gauge field $h^{a}$ to the Rarita-Schwinger and Einstein-Hilbert terms. In turn, the $\tilde{C}_{4}^{\textcircled{\footnotesize 1}}$ superalgebra assures the pure supergravity action without cosmological constant ($\alpha _{1}$ sector), while the extra field $h^{a}$ appears only in the $\alpha _{2}$ sector. Interestingly, only the Soroka-Soroka superalgebra offers the possibility to construct a supergravity action with the cosmological constant. On the other hand, although there are well-defined supersymmetric extensions of the $\tilde{B}_{4}^{\textcircled{\footnotesize 2}}$, $\tilde{C}_{4}^{\textcircled{\footnotesize 2}}$ and  $\mathfrak{B}_{4}$ algebras, such superalgebras reproduce non-standard supersymmetric CS actions with no fermionic contributions in the $\alpha _{1}$ term. In the case of the $B_{4}$ algebra there is no consistent supersymmetric extension with one fermionic charge. This gives us a purely bosonic CS action, which we left in the table for comparison.

As an ending remark, naturally one could extend our results to four spacetime dimensions. This could be achieved by using the MacDowell-Mansouri formalism \cite{MacDowell:1977jt}. In particular, the geometrical formulation of supergravity in such formalism has already been presented for the Maxwell and Soroka-Soroka superalgebra in \cite{Concha:2014tca, Concha:2015tla, Durka:2011gm}. Then, one could generalize the procedure to new resonant superalgebras containing Poincaré-like subalgebras. However, unlike the three-dimensional cases, all the new resonant superalgebras would lead to the actions missing both Einstein-Hilbert and the cosmological constant. Nevertheless, it would be interesting to analyze the requirements for the additional gauge field $h_a$ in order to construct dynamical actions.

\subsection{Flat limit and Inönü-Wigner contraction}

As we have shown previously, the supersymmetric action based on the non-standard Maxwell superalgebra $\mathfrak{B}_4$ can be obtained as a vanishing cosmological constant limit $\ell \rightarrow \infty$ of the $\mathfrak{C}_4$ supergravity theory. Such flat limit can be applied explicitly at the level of the (anti-)commutation relations \eqref{SorokaExplicit}, the invariant tensor \eqref{C4inv}, the curvature two-forms \eqref{curvs} and CS action \eqref{SSCS}. It is intriguing to see that the vanishing of the cosmological constant leads us to the non-standard supersymmetric theory whose action describes no more a supergravity theory. As we have previously discussed, such peculiarity can be avoided by considering an additional fermionic generator leading to the so-called minimal scheme. Configurations with more than one spinor charge will be the subject of some future work.

As we shall see, the flat limit is not the only limit allowing us to relate two resonant superalgebras. Indeed, various rescalings of the generators and the arbitrary constant appearing in the invariant tensor allow establishing a well-defined IW contraction at the level of the superalgebra and CS action. To this end, we consider a rescaling parameter $\sigma$ and apply the limit $\sigma \rightarrow \infty$. The table below contains the different rescalings of the generators and coupling constants appearing in the $\{\mathfrak{C}_{4},\tilde{C}_{4}^{\textcircled{\footnotesize i}}\}$ family which in the limit $\sigma\rightarrow\infty$ reproduce the $\{\mathfrak{B}_4,\tilde{B}_{4}^{\textcircled{\footnotesize i}}\}$ superalgebras with their respective invariant tensors.

\begin{equation*}
\begin{tabular}{l|c|c|c|c}
& $\mathfrak{C}_{4}\rightarrow\mathfrak{B}_{4}$ & $\mathfrak{C}_{4}\rightarrow \tilde{B}_{4}^{\textcircled{\footnotesize 3}}$ & $\tilde{C}_{4}^{\textcircled{\footnotesize 1}}\rightarrow \tilde{B}_{4}^{\textcircled{\footnotesize 1}}$ & $\tilde{C}_{4}^{\textcircled{\footnotesize 2}}\rightarrow\tilde{B}_{4}^{\textcircled{\footnotesize 2}}$ \\ \hline
$J_{a}$ & $J_{a}\rightarrow J_{a}$ & $J_{a}\rightarrow J_{a}$ & $J_{a}\rightarrow J_{a}$ & $J_{a}\rightarrow J_{a}$ \\
$P_{a}$ & $P_{a}\rightarrow \sigma P_{a}$ & $P_{a}\rightarrow \sigma^{2} P_{a}$ & $P_{a}\rightarrow P_{a}$ & $P_{a}\rightarrow P_{a}$ \\
$Z_{a}$ & $Z_{a}\rightarrow \sigma^{2} Z_{a}$ & $Z_{a}\rightarrow \sigma^{2} Z_{a} $ & $Z_{a}\rightarrow \sigma Z_{a} $ & $Z_{a}\rightarrow \sigma^{2} Z_{a} $ \\
$Q_{\alpha}$ & $Q_{\alpha} \rightarrow \sigma Q_{\alpha}$ & $Q_{\alpha} \rightarrow \sigma Q_{\alpha}$ & $Q_{\alpha} \rightarrow Q_{\alpha}$ & $Q_{\alpha} \rightarrow \sigma Q_{\alpha}$ \\ \hline
$\alpha_0$ & $\alpha_0\rightarrow \alpha_{0}$ & $\alpha_0\rightarrow \alpha_{0}$ & $\alpha_0\rightarrow \alpha_{0}$ & $\alpha_0\rightarrow \alpha_{0}$ \\
$\alpha_1$ & $\alpha_1 \rightarrow \sigma \alpha_1$ & $\alpha_1 \rightarrow \sigma^{2} \alpha_1$ &$\alpha_1 \rightarrow \alpha_1$ & $\alpha_1 \rightarrow \alpha_1$ \\
$\alpha_{2}$ & $\alpha_{2}\rightarrow\sigma^{2}\alpha_{2}$ & $\alpha_{2}\rightarrow\sigma^{2}\alpha_{2}$ & $\alpha_{2}\rightarrow\sigma\alpha_{2}$ & $\alpha_{2}\rightarrow\sigma^{2}\alpha_{2}$
\end{tabular}%
\end{equation*}%

Let us note that no limits are explored for the $B_{4}$ algebra since there is no valid supersymmetric extension of such symmetry. On the other hand, one can see that the vanishing cosmological constant limit, relating $\mathfrak{C}_{4}$ and $\mathfrak{B}_{4}$ superalgebras, can be interpreted as an IW contraction process analogously to what happens in the AdS and Poincaré superalgebras. Although the other resonant superalgebras related by an IW contraction do not need to rescale the constants of the invariant tensor, one can see that the proper contraction of the CS action requires appropriate rescaling of the $\alpha$'s. It is interesting to point out that the presence of an additional bosonic generator concerning the AdS and Poincaré symmetries not only offers diverse ways to close a superalgebra but also leads us to several relations between the superalgebras through the deformation/contraction procedure.

\section{Conclusions}

In this paper, we have analyzed all possible $\mathcal{N}=1$ supersymmetric extensions of the so-called resonant algebras \cite{Durka:2016eun, Durka:2019guk, Durka:2019vnb} by studying the super-Jacobi identities. To this purpose, we provided a list of minimal requirements for the superalgebras. We obtained not only the already known Maxwell and Soroka-Soroka superalgebras \cite{Soroka:2006aj,Lukierski:2010dy} but also novel superalgebras, which possess a Poincaré-like bosonic structure. Interestingly, some particular resonant algebras offer more than one supersymmetric extension. Altogether, for the $JPZ+Q$ configuration, we have discovered eight different cases. However, as we have discussed, only five supersymmetric extensions of the resonant algebras reproduce consistent CS supergravity actions, whereas the other three represent non-standard examples.

Gathering all these superalgebras in one place with their respective CS actions, immediately highlights many relations, differences, and similarities between them. It is interesting to see how subtle differences in commutation relations can lead to important differences in the CS action. In particular, regarding only the Einstein-Hilbert term within $\alpha_1$ sector, the five "good" resonant superalgebras allow us to describe diverse supergravity CS actions with and without cosmological constant and the contribution from the extra gauge field $h_{a}$. It would be worthwhile to explore full physical implications of such an extra field. Although there are some interpretations of an additional $h^{a}$ gauge field in the three-dimensional Maxwell CS gravity theory \cite{Concha:2018zeb, Bansal:2018qyz, Chernyavsky:2020fqs, Durka:2019guk}, the fermionic framework remains still an opened issue.

Extending our results to more fermionic charges seems a natural next step. Due to the presence of a second spinorial charge, one can expect a greater number of possibilities. Let us note that the presence of a second spinorial charge is not new and has been previously introduced in superstring context \cite{Green:1989nn} and $D=11$ supergravity theory \cite{DAuria:1982uck}. Such (minimal) supersymmetric extensions of the $\mathfrak{B}_{4}$ and $\mathfrak{C}_{4}$, without considering additional bosonic content, have been recently obtained as the resonant expansions of the Lorentz superalgebra \cite{Concha:2018jxx, Concha:2019icz}. As it was discussed in \cite{Concha:2018jxx, Concha:2019icz}, unlike the non-standard case, a minimal supersymmetric extension of the Maxwell algebra $\mathfrak{B}_{4}$ allows defining a consistent CS supergravity action in three spacetime dimensions. Then, It would be also interesting to explore supersymmetric extensions of other resonant algebras with more fermionic charges and bosonic content.

Finally, our results might provide valuable information about the underlying symmetry behind supergravity in higher dimensions. It would be interesting to explore the possibility to recover Cremmer-Julia-Scherk (CJS) supergravity \cite{Cremmer:1978km} from a more general CS supergravity based on enlarged superalgebras.

\section{Acknowledgments}

This work was funded by the National Agency for Research and Development (ANID) CONICYT - PAI grant No. 77190078 (P.C.), the FONDECYT Project N$^{\circ }$3170438 (E.R.) and the Institute Grant for Young Researchers 0420/2716/18 (R.D.). P.C. would like to thank to the Dirección de Investigación and Vice-rectoría de Investigación of the
Universidad Católica de la Santísima Concepción, Chile, for their constant support.

\end{document}